\def\hb{H$\beta$}
\def\mgii{Mg {\sc ii}~}
\def\civ{C {\sc iv}~}

\documentclass[referee,usenatbib]{mn2e}
\usepackage{aas_macros}
\usepackage{subfigure}
\usepackage{graphicx,natbib,rotating}

\title[The Compact Structure of Radio-Loud Broad Absorption Line Quasars]
{The Compact Structure of Radio-Loud Broad Absorption Line
Quasars}
\author[Liu et al.]
{Y. Liu $^{1,3}$\thanks{E-mail: yliu@shao.ac.cn}, D. R.
Jiang$^{1,3}$, T. G. Wang$^{2,3}$, F. G. Xie$^{1,3,4}$
\\
$^{1}$ Shanghai Astronomical Observatory, Chinese Academy of
Sciences, Shanghai 200030, China.
\\
$^{2}$Center for Astrophysics, University of Science and
Technology of China, Hefei, Anhui 230026, China
\\
$^{3}$Joint Institute for Galaxy and Cosmology (JOINGC) of SHAO
and USTC
\\
$^{4}$Graduate School of Chinese Academy of Sciences, Beijing
100039, China}

\begin{document}
\pagerange{\pageref{firstpage}--\pageref{lastpage}} \pubyear{2002}
\maketitle
\label{firstpage}
\begin{abstract}

We present the results of EVN+MERLIN VLBI polarization
observations of 8 Broad Absorption Line (BAL) quasars at 1.6 GHz,
including 4 LoBALs and 4 HiBALs with either steep or flat spectra
on VLA scales. Only one steep-spectrum source, J1122+3124, shows
two-sided structure on the scale of 2~kpc. The other four
steep-spectrum sources and three flat-spectrum sources display
either an unresolved image or a core-jet structure on scales of
less than three hundred parsecs. In all cases the marginally
resolved core is the dominant radio component. Linear polarization
in the cores has been detected in the range of a few to 10
percent. Polarization, together with high brightness temperatures
(from 2$\times10^9-5\times10^{10}$K), suggest a synchrotron origin
for the radio emission. There is no apparent difference in the
radio morphologies or polarization between low-ionization and
high-ionization BAL QSOs nor between flat- and steep-spectrum
sources. We discuss the orientation of BAL QSOs with both flat and
steep spectra, and consider a possible evolutionary scenario for
BAL QSOs. In this scenario, BAL QSOs are probably the young
population of radio sources, which are Compact Steep Spectrum or
GHz peaked radio source analog at the low end of radio power.

\end{abstract}


\section{Introduction}
About~$10$~-~$20$\% of optically selected quasars exhibit Broad
Absorption Line (BAL) troughs up to $\sim 0.1c$ blueward of their
corresponding emission lines (Weymann et al. 1991; Weymann 2002;
Tolea et al. 2002; Hewett \& Foltz 2003; Reichard et al. 2003).
These BALs are produced via resonant scattering in partially
ionized outflows. BAL quasars can be further divided into high-
and low-ionization sub-classes (HiBALs and LoBALs). HiBAL quasars
are those objects which show BALs only in highly ionized species
such as CIV and NV, while LoBAL quasars also show BALs in Mg II or
Al III. Two different scenarios have been proposed to explain the
BAL phenomena. In the first scenario, the BAL region (BALR) is
present with a small covering factor in every QSO. The different
appearance of BAL and non-BAL QSOs is solely due to different
lines of sight. In a BAL QSO our line of sight intercepts the BALR
while it does not in non-BAL QSOs. In this scenario, the fraction
of BAL QSOs is interpreted as the covering factor of BALRs. This
paradigm has also been generalized to interpret the dichotomy of
LoBAL and HiBAL as an orientation effect, in which low-ionization
outflow covers small fraction sky with the BAL outflow. This
scenario is consistent with a number of statistical properties,
such as the similarity between the UV continuum and emission line
spectra for both types of QSOs (e.g., Weymann et al. 1991), their
similar millimeter and far-infrared luminosities (e.g., Willott et
al. 2003), as well as their similar large scale environments (Shen
et al. 2008). In the second scenario BALs exist only in a relative
short phase of QSO activity, which is very likely in their early
phase of evolution, at least for LoBAL QSOs (Briggs, Turnshek \&
Wolfe 1984). Support for the latter scenario includes the excess
fraction of LoBAL quasars among infrared selected quasars (Boroson
\& Meyers 1992), and a special locus in black hole mass versus
accretion rate space (Boroson 2002).

In the first scenario the BALR is probably located in a preferred
direction with respect to the system$^{\rm '}$s symmetry axis,
e.g., in the equatorial or polar direction. Higher polarization in
the optical continuum in comparison to non-BAL QSOs suggests that
BAL QSOs are seen nearly edge-on (Goodrich \& Miller 1995; Hines
\& Wills 1995; Cohen et al. 1995; Wang, Wang \& Wang 2005).
Equatorial outflows are also preferred from theoretical
considerations for radiatively accelerated winds from an accretion
disc or evaporated gas from a putative dusty torus (Punsly 2006).

However, there are indications for both polar and equatorial
outflows in radio-loud BAL QSOs.  It is generally believed that
the radio jet is aligned with the symmetry axis of the accretion
disk. Thus it can be taken as an indicator of the system
inclination. According to the unification scheme of radio quasars,
a radio-loud quasar would show a flat radio spectrum and a
core-dominated morphology when viewed along the radio jet, due to
relativistic boosting of the optically thick core. It would appear
as a steep-spectrum radio source with double-lobed morphology when
viewed side-ways (e.g., Urry \& Padovani 1995). Thus the radio
morphology and spectrum can serve as a surrogate for the
inclination of the system: in an equatorial outflow model a
radio-loud BAL QSO would appear as a steep-spectrum radio source
with double-lobed radio morphology, while the opposite situation
will be observed in a polar-outflow model. Becker et al. (2000)
found that BAL QSOs in the FIRST Bright Quasar Survey (FBQS)
display both steep and flat radio spectra (see also Menou et al.
2001), indicating both polar and equatorial outflows. Other
studies suggest a similar result. On the one hand, nearly a dozen
double-lobed FR~II quasars were found, mostly from the SDSS and
FIRST samples (Wills, Brandt, \& Loar 1999; Gregg et al. 2000;
Brotherton et al. 2002; Zhou et al. 2006; Gregg, Becker \& de
Vries 2006). On the other hand, evidence for polar outflows has
also been found in a small number of radio variable BAL QSOs (Zhou
et al 2006; Ghosh \& Punsly 2007), in which the radio variability
suggests that the radio jet is beamed towards the observer. The
presence of both polar and equatorial outflows can be better
interpreted in the context of an evolutionary scenario, rather
than the geometrical unification scheme.

Further studies of radio-loud BAL QSOs are necessary in order to
fully understand the geometry of outflows in their population.
Note that most radio-loud BAL QSOs with steep radio spectra show
compact structures on VLA scales, different from classical
radio-loud quasars on which the unification scheme is based. Thus
it is not clear whether the radio spectrum can be used as an
indicator for inclination or not. Powerful FR~II radio BAL QSOs
are extremely rare among radio-loud BAL QSOs, and most radio BAL
QSOs are only moderately strong in the radio. Morphological
studies of these radio intermediate BAL quasars can reveal
representative properties of the majority radio population of BAL
QSOs. Currently, less than a handful of such BAL QSOs have been
studied at sufficiently high angular resolution and they have
revealed a complex situation.

In previous EVN observations, Jiang \& Wang (2003) observed 3 BAL
quasars at L band using the phase reference technique. Among these
three sources, only one (J1312+2319) has compact two-sided
structure, and the other two are still unresolved. The jet
orientation might be near the line of sight if the compact
component is the jet base in these two unresolved sources. In this
case the orientation model may be problematic, at least for these
3 sources.

In this paper, we report our EVN+MERLIN observations at 1.6 GHz of
eight BAL quasars, including equal numbers of HiBAL and LoBAL
quasars and then give some discussion of the results. $\S$ 2
presents the observations as well as data reduction and the
results are described in $\S$ 3. The discussion is contained in
$\S$ 4, and in the last section $\S$ 5 we give the conclusions.
Throughout the paper we adopt the spectral index convention $\rm
f_{\nu}\propto\nu^{-\alpha}$ and a cosmology with $\rm H_{0}=70
\rm {~km ~s^ {-1}~Mpc^{-1}}$, $\rm \Omega_{M}=0.3$, $\rm
\Omega_{\Lambda} = 0.7$. All values of luminosity used in this
paper are calculated with our adopted cosmological parameters.

\section{Observation and Data Reduction}
Our BAL quasars are comparatively bright sources~($\sim$ 10 mJy),
which have been selected from the Becker et al. (2000) BAL quasar
sample. EVN+MERLIN phase-reference polarimetric continuum
observations were made for 8 BAL quasars at L band on 2005 June
22. The observations were divided into two runs, each lasting 12
hours. In order to compare the radio morphology of either low- and
high-ionization or steep- and flat-spectrum BAL quasars, our
sample included 4 LoBAL quasars and 4 HiBAL quasars with both
steep and flat spectra on VLA scales. The left/right-circular
polarization signals were recorded in 8 baseband channels with a
total bandwidth of 32 MHz and 2-bit sampling. The angular
distances of the phase referencing sources to the target sources
are less than $\rm 3^{\circ}$. The scan time for each source was
about 1.5 hours, and the estimated thermal noise in the total
intensity images was no more than 0.1 mJy/beam. We used
OQ208/J1407+284 and J0927+3902 as the D-term calibrators for the
1st and 2nd runs, respectively, and 3C286/J1331+305 as the
electric vector
polarization angle~(EVPA) calibrator for both runs. 
The EVN data were correlated at the Joint Institute
for VLBI in European~(JIVE) in Dwingeloo. 
Pipeline results were used for the phase referencing. Since the
flux densities of all BAL quasars in our sample except J1603+3002
are lower than 20 mJy on VLA scales, we used non-phase referencing
results of J1603+3002~(about 54 mJy, which is strong enough to
detect with normal fringe fitting)~to compare with the results
from phase referencing.
The results 
are in good agreement, suggesting that the phase referencing
observations of the other sources should also be reliable. The
initial amplitude calibration and the fringe-fitting were
performed using the NRAO Astronomical Image Processing System
(AIPS). The imaging and self-calibration were carried out in the
DIFMAP package (Shepherd, Pearson, \& Taylor 1994) and these
self-calibrations allowed us to improve the signal to noise ratio
in the images. The estimated uncertainty of the amplitude
calibration is about 10 percent. The detailed information for
these observations are listed in Table 1 and
Table 2, 
including total flux densities from both EVN and MERLIN, positions
shifts, fractional polarization, EVPA as well as uncertainty for
polarization. Moreover, making use of the phase-reference
technique, we were able to obtain positions from the VLBI data
which are more accurate
than those obtained from the VLA. Table 3 lists all the position
information for target BAL sources and phase-reference sources.
The position of the phase tracking center for BAL sources is shown
in column (2) and column (3), while the name and position of the
phase-reference sources are denoted in column (4), column (5) and
column (6), respectively. In our sample, J1044+3656 has the
largest shift away from its phase tracking center, about 225 mas
east and 225 mas south. Although these shifts might affect the
signal to noise ratio, the EVN flux densities of our sources are
similar to those observed at the VLA, suggesting that their
influence is too small to be important. The accurate positions of
the BAL quasars calculated from the shifts away from the phase
tracking centers are listed in the last two columns in Table 3.
The 3$\rm \sigma$ uncertainty in these positions, which mainly
depends on both the angular resolution of the EVN at 18 cm and the
signal to noise ratio, is $\rm \sim$15 mas.

\begin{table*} \caption{{\large Sources observed with EVN+MERLIN
at 18 cm.} The source name (in J2000) and GST range are shown in
Column(1) and Column(2). The VLA spectral index and VLA flux
density at 20 cm are listed in Column(3) and Column(4), while the
EVN and MERLIN flux density at 18 cm are shown in Column(5) and
Column(6) respectively. The last Column is the classification of
BAL quasars.}
\begin{tabular}{lccccccccccc}
\hline\hline Name(s) & GST range  & $\alpha$ & VLA & EVN  & MERLIN & Note \\
  & (Europe)   & ($S_\nu\propto \nu^{\alpha}$) & (mJy at 20 cm) & (mJy at 18 cm) &(mJy at 18 cm) & & \\
\hline
$ J0724+4159$ & 22:00-14:00   & 0.0   &  7.9~(B) &  7.2 &  8.0 & LoBAL \\
$ J0728+4026$ & 22:00-14:00   & -1.1  & 18.0~(A) & 16.1 & 17.6 & LoBAL \\
$ J1044+3656$ & 02:00-18:00   & -0.5  & 15.6~(A) & 16.4 & 18.5 & LoBAL \\
$ J1122+3124$ & 03:00-17:00   & -0.6  & 12.9~(B) &  8.3 & 10.2 & LoBAL \\
$ J1150+2819$ & 04:00-17:00   & -1.2  & 14.2~(B) & 12.6 & 12.3 & HiBAL \\
$ J1413+4212$ & 05:00-21:00   & -0.2  & 18.7~(B) & 17.1 & 18.1 & HiBAL \\
$ J1603+3002$ & 08:00-21:00   & -0.6  & 54.2~(A) & 52.1 & 52.6 & HiBAL \\
$ J1655+3945$ & 08:00-23:00   & -0.2  & 10.2~(A) & 12.1 & 10.6 & HiBAL \\
\hline
\end{tabular}
\begin{quote}
\ Note: GST range is unique, eliminate reference to Europe.
\end{quote}
\end{table*}

\begin{table*}
\caption{{\large EVN+MERLIN observational details of BAL quasars
at 18 cm.}}
\begin{tabular}{lccccccccccc}
\hline\hline Sources Name & Position & Position & m & $\rm \sigma_{m}$ & EVPA & $\rm \sigma_{EVPA}$  \\
                      & EVN (mas)& MERLIN (mas) & ($\%$ in EVN) & ($\%$)& ($^{\circ}$)&($^{\circ}$)  \\
\hline
$ J0724+4159$   & (-129, 228) & (-150, 210) & 11   & 3.9  &  113.8  & 11.6 \\ 
$ J0728+4026$   & (-126, 224) & (-120, 210) & 4.7  & 1.5  &  131.9  & 34.7 \\ 
$ J1044+3656$   & (225, -225) & (210, -240) & 3.2  & 1.3  &  125.9  & 32.9 \\ 
$ J1122+3124$   & (-123, -200)& (-120, -210)& 3.0  & 1.3  &  81.9   &  7.7 \\ 
$ J1150+2819$   &   (45, -58) &  (30, -60)  & 3.1  & 1.3  &   5.8   & 30.0 \\ 
$ J1413+4212$   & (-158, -5)  & (-180, 0)   & 3.3  & 1.3  &   85.5  & 32.9 \\ 
$ J1603+3002$   & (-104, -171)& (-120, -180)& 1.3  & 0.3  &  126.4  & 34.7 \\ 
$ J1655+3945$   & (18, -66)   & (30, -60)   & 4.1  & 1.4  &   49.2  & 26.6 \\ 
\hline
\end{tabular}
\begin{quote}
\ Column(1): IAU name in J2000; Column(2): distance from EVN phase
tracking center; Column(3): distance from MERLIN phase tracking
center; Column(4): percentage polarization in EVN observations;
Column(5): uncertainty for percentage polarization in EVN
observations; Column(6): EVPA in EVN observations; Column(7):
uncertainty for EVPA.
\end{quote}
\end{table*}

\begin{table*}
\caption{{\large The VLBI positions of 11 BAL quasars, including 3
previously observed BAL quasars.}} \center
\begin{tabular}{lccccccccccc}
\hline\hline Source & RA (hh mm) & Dec (dd mm) & reference & reference RA & reference Dec & RA (hh mm) & Dec (dd mm) \\
                    & (J2000)    & (J2000)     &           &(J2000)     &  (J2000)        &  (J2000)       & (J2000)\\
\hline
$ J0724+4159$     & 07 24 18.4920   & +41 59 14.400 & $J0730+4049$ &07 30 51.3466 &+40 49 50.827 &07 24 18.4834 & +41 59 14.628\\ 
$ J0728+4026$     & 07 28 31.6610   & +40 26 15.850 & $J0730+4049$ &07 30 51.3466 &+40 49 50.827 &07 28 31.6526 & +40 26 16.074\\ 
$ J1044+3656$     & 10 44 59.5910   & +36 56 05.390 & $J1050+3430$ &10 50 58.1230 &+34 30 10.941 &10 44 59.6060 & +36 56 05.165 \\ 
$ J1122+3124$     & 11 22 20.4620   & +31 24 41.200 & $J1130+3031$ &11 30 42.4292 &+30 31 35.388 &11 22 20.4538 & +31 24 41.000\\ 
$ J1150+2819$     & 11 50 23.5700   & +28 19 07.500 & $J1147+2635$ &11 47 59.7639 &+26 35 42.333 &11 50 23.5730 & +28 19 07.442\\ 
$ J1413+4212$     & 14 13 34.4040   & +42 12 01.760 & $J1405+4056$ &14 05 07.7954 &+40 56 57.831 &14 13 34.3935 & +42 12 01.755\\ 
$ J1603+3002$     & 16 03 54.1620   & +30 02 08.880 & $J1605+3001$ &16 05 33.0480 &+30 01 29.702 &16 03 54.1551 & +30 02 08.709\\ 
$ J1655+3945$     & 16 55 43.2350   & +39 45 19.940 & $J1652+3902$ &16 52 58.5096 &+39 02 49.823 &16 55 43.2362 & +39 45 19.874\\ 
$ J0957+2356^{*}$ & 09 57 07.3670   & +23 56 25.320 & $J0956+2515$ &09 56 49.8754 &+25 15 16.050 &09 57 07.3712 & +23 56 25.379\\
$ J1312+2319^{*}$ & 13 12 13.5600   & +23 19 58.510 & $J1321+2216$ &13 21 11.2014 &+22 16 12.092 &13 12 13.5753 & +23 19 58.572 \\
$ J1556+3517^{*}$ & 15 56 33.7720   & +35 17 57.620 & $J1602+3326$ &16 02 07.2635 &+33 26 53.072 &15 56 33.7768 & +35 17 57.389 \\
 \hline
\end{tabular}
\begin{quote}
\ Column(1): IAU name of BAL quasar in J2000, $*$~denotes
perviously observed source; Column(2): RA measured by VLA in
J2000; Column(3): Dec measured by VLA in J2000; Column(4): IAU
name of phase-reference source in J2000; Column(5): RA for
phase-reference source; Column(6): Dec for phase-reference source;
Column(7): RA remeasured for BAL quasar by our EVN observation in
J2000; Column(8): Dec remeasured for BAL quasar by our EVN
observation in J2000.
\end{quote}
\end{table*}

\begin{table*}
\caption{{\large Optical line information from SDSS released data
and black hole mass for sources in our sample.}}
\begin{tabular}{lccccc}
\hline\hline ~~~~~Sources Name & Redshift   & $f_{\lambda3000}$  & FWHM  & lines & $M_{\rm BH}$ \\
          &   & ($\rm 10^{-17}~ergs~s^{-1}\AA^{-1}cm^{-2}$)& ($\rm km~s^{-1}$) & & $10^{8}~\rm M_{\odot}$  \\
\hline
~~~~~$J0724+4159$~~~~~ & ~~~~~1.551~~~~~ & ~~~~~   85~~~~~ & ~~~~~ 2428~~~~~ & ~~~~~ \mgii~~~~~ & 0.26 \\ 
~~~~~$J1044+3656$~~~~~ & ~~~~~0.702~~~~~ & ~~~~~  201~~~~~ & ~~~~~ 4197~~~~~ & ~~~~~ \hb  ~~~~~ & 3.98 \\ 
~~~~~$J1122+3124$~~~~~ & ~~~~~1.453~~~~~ & ~~~~~  124~~~~~ & ~~~~~ 2518~~~~~ & ~~~~~ \mgii~~~~~ & 0.34 \\ 
~~~~~$J1150+2819$~~~~~ & ~~~~~3.124~~~~~ & ~~~~~  323~~~~~ & ~~~~~ 2069~~~~~ & ~~~~~ \civ ~~~~~ & 0.59 \\ 
~~~~~$J1413+4212$~~~~~ & ~~~~~2.817~~~~~ & ~~~~~ 225 ~~~~~ & ~~~~~ 6116~~~~~ & ~~~~~ \civ ~~~~~ & 4.47 \\ 
~~~~~$J1603+3002$~~~~~ & ~~~~~2.030~~~~~ & ~~~~~  95 ~~~~~ & ~~~~~ 1619~~~~~ & ~~~~~ \mgii~~~~~ & 0.11 \\ 
~~~~~$J1655+3945$~~~~~ & ~~~~~1.753~~~~~ & ~~~~~ 80  ~~~~  & ~~~~~ 1529~~~~~ & ~~~~~ \mgii~~~~~ & 0.09 \\ 
\hline
\end{tabular}
\begin{quote}
\ Col. (1): Object name. Col. (2): Redshift. Col. (3): flux
density at 3000$\AA$ in units of $\rm
10^{-17}~ergs~s^{-1}\AA^{-1}cm^{-2}$. Col. (4): full width at half
maximum (FWHM) of broad line in units of $\rm km~s^{-1}$. (5):
line used for FWHM. Col. (6): black hole mass in unit of
$10^{8}~\rm M_{\odot}$. See Kaspi et al. (2005) and McLure \&
Jarvis (2002) for the detailed method of the black hole mass
calculation.
\end{quote}
\end{table*}

\section{Sources and Results}

\subsection{J0724+4159}
J0724+4159 is a LoBAL quasar. The radio source associated with
this quasar is unresolved in the VLA observation at 20 cm. Its
flux density is 7.9 mJy in FIRST. J0724+4159 has a spectral index
between 20 cm and 3.6 cm of $\alpha = 0.0$. The radio loudness is
$logR=1.6$~(Becker et al. 2000). The naturally weighted image with
the EVN (Fig.1(a)) still shows an unresolved structure at 18 cm,
with a total flux density of 7.2 mJy. With the same weighting, the
image derived from MERLIN gives a total flux density of 8.0 mJy.
Since the difference in the flux density between EVN and MERLIN is
consistent with the uncertainty of the flux calibration, we
believe that there are no missing components in J0724+4159 at high
angular resolution with EVN. The compact component in the EVN
image of J0724+4159 shows a high polarization level of $\sim$
11\%.

\subsection{J0728+4026}
J0728+4026 was discovered as a LoBAL quasar. It is a point source
with 18 mJy with the VLA at 18 cm~(A-configuration). The flux
density with the VLA A-configuration at 3.6 cm is 2.7 mJy. These
non-simultaneous data suggest a steep spectral index $\alpha =
-1.1$. Its radio loudness is $logR=0.37$~(Becker et al. 2000). The
naturally weighted image with the EVN (Fig.1(b)) shows a core-jet
structure. The flux density in the core-jet structure is 16.1 mJy
in EVN observation, which accounts for about 92\% of the total
flux density of 17.3 mJy detected by MERLIN. The core-jet
structure has a degree of polarization of about 4.7\%.

\subsection{J1044+3656}
J1044+3656 is a LoBAL quasar. The VLA observation in
A-configuration at 20 cm suggests that this source is unresolved
with a flux density of about 15.6 mJy. The spectral index between
20 cm and 3.6 cm is $\alpha = -0.5$. The radio loudness is
$logR=1.29$~(Becker et al. 2000). The naturally weighted image
with the EVN (Fig.1(c)) displays a core-jet structure at 18 cm
with a total flux density of 16.4 mJy. MERLIN detected a flux
density of about 18.5 mJy. There might be some weak structures
resolved out in the EVN image. The difference in the flux density
between the VLA and MERLIN images is probably due to source
variation.

J1044+3656 shows a fractional polarization of about 3.2\% in the
central component. The two different directions of the electric
vector in the compact region in the EVN map may come from two
different components, one being the core and another the jet
component. The jet component might be only just separated from the
core, its direction of electric vector following the jet
direction.

\subsection{J1122+3124}
J1122+3124 is a LoBAL quasar. The spectral index between 20 cm and
3.6 cm obtained by the VLA in D-configuration is $\alpha = -0.6$.
The estimated radio loudness is $logR=1.52$~(Becker et al. 2000).
The naturally weighted image of the source with the EVN (Fig.1(d))
shows a central component with two-sided structure at 18 cm, which
is nearly aligned with the central component. The southern
component is located 150 mas away from the central component,
while a weak northern component lies 87 mas away. About 90\% of
the flux density is present in the central component. In order to
confirm the two-sided structure, we combined EVN and MERLIN data
with different weights~(see Fig.2 ). It seems that the northern
component has more extended structure. The flux densities of the
clean components in the EVN and MERLIN images are 8.3 mJy and 10.2
mJy, respectively. The extended northern emission was probably
resolved at EVN resolution. The source may vary. The flux density
measured by the VLA is 12.9 mJy in B-configuration at 20 cm, and
another two measurements at the same frequency with the VLA are
10.9 mJy and 10.0 mJy in D-configuration~(Becker et al. 2000).

The naturally weighted EVN image shows about 3.0\% fractional
polarization in the central component, but it is not detected in
the two-sided lobes because of their weak intensity. The
fractional polarization with MERLIN is about 2.7\%, which is
consistent with the EVN result.

\subsection{J1150+2819}
J1150+2819 is a HiBAL quasar. The VLA observation at 20 cm shows
an unresolved image with a flux density of about 14.2 mJy. The
spectral index between 20 cm and 3.6 cm is $\alpha = -1.2$. The
estimated radio loudness is $logR=1.43$~(Becker et al. 2000). The
naturally weighted image with the EVN (Fig.3(a)) still displays
unresolved structure at 18 cm. The clean components sum to 12.6
mJy in the EVN measurement, and 12.3 mJy with MERLIN. J1150+2819
shows a fractional polarization of about 3.1\% in the central
component.

\subsection{J1413+4212}
J1413+4212 is a HiBAL quasar. It is a point source with 18.7 mJy
in the FIRST catalog. The flux density measured with the VLA at
3.6 cm in the D-configuration is 11.3 mJy. This gives a flat
spectral index $\alpha = -0.2$. The radio loudness is
$logR=1.68$~(Becker et al. 2000). The naturally weighted image
with the EVN (Fig.3(b)) displays a short core-jet structure at 18
cm with 17.1 mJy. The MERLIN array measures about 18.1 mJy in
clean components. The central component has 3.3\% fractional
polarization.

\subsection{J1603+3002}
J1603+3002 is a HiBAL quasar. It is the most luminous source in
these observations. It has a radio loudness $logR=2.04$~(Becker et
al. 2000). The flux densities at 20 cm and 3.6 cm detected by the
VLA in A-configuration are 54.2 mJy and 18.1 mJy respectively,
which suggests a spectral index of $\alpha = -0.6$. The total flux
density is about 52.1 mJy in the EVN image~(see Fig.3(c)). The
MERLIN image shows an unresolved structure with 52.6 mJy flux
density. The fractional polarization of the EVN image is about
1.3\%.

\subsection{J1655+3945}
J1655+3945 is a HiBAL quasar. The VLA observation at 20 cm shows
an unresolved image. The flux density of J1655+3945 in the FIRST
catalog is 10.2 mJy. The spectral index between 20 cm and 3.6 cm
is $\alpha = -0.2$. Its radio loudness is $logR=1.41$~(Becker et
al. 2000). The naturally weighted EVN image depicts a resolved
structure elongated to the south west with a total of 12.1
mJy~(Fig.3(d)). 
In the central component the fractional polarization is about
4.1\%.

\section{Discussion}

The origin of the radio emission from these radio intermediate
quasars is not fully understood. Based on the high brightness
temperature inferred from radio variability, Zhou et al. (2006)
suggested that the emission comes from relativistic jets for a
small sub-sample of radio-loud BAL QSOs, while Blundell \& Kuncic
(2007) argued that the emission is produced by free-free emission
from outflows in radio-weak QSOs. The brightness temperature in
the source rest frame for the strongest components of these BAL
quasars, based on the results of model-fitting, range from
$2.0\times10^{9}$K to $5.2\times10^{10}$K. The high brightness
temperature and moderate polarization degree suggest that the
radio is synchrotron emission from the jet.

Including the two-sided structure source J1122+3124, all BAL
quasars that we observed with EVN+MERLIN at 18 cm exhibit a
compact structure with a projected size of less than 2 kpc. In
addition to J1122+3124, the total flux densities of the other
seven compact sources measured with EVN at 18 cm are very close to
the flux densities measured by both our simultaneous MERLIN and
previous VLA observations at 20 cm. Therefore, if there is any low
brightness extended structure in these object, it must be weak.
Assuming that the 3.6 cm radio emission of these BAL quasars
measured with the VLA comes from the compact component and they
are non-variable, their radio spectra can be calculated by using
VLA measurements between 20 cm and 3.6 cm. Five are steep-spectrum
radio sources, and the rest are flat-spectrum sources. Combined
with our previous observations of another three compact BAL
quasars~(see $\S$ 1)~(Jiang \& Wang 2003), there are seven
steep-spectrum BAL quasars and four flat-spectrum sources. In
these 11 compact BAL quasars observed with the EVN~(or EVN+MERLIN)
array, there is no significant difference between HiBAL and LoBAL
quasars in their radio morphology and polarization. Similarly,
there is no apparent difference in radio properties between flat-
and steep-spectrum BAL quasars.

Among the seven steep-spectrum sources, five (J0728+4026,
J1044+3656, J1150+2819, J1603+3002 and J0957+2356) show only
compact cores or core-jet structures. In these cases there is no
good indicator for the symmetry axis of the system, and the
inclination might be distributed over a large range of values.
However, although their sizes are much smaller than the typical
values for Compact Steep Spectrum (CSS) sources ($\sim$ 15 kpc),
it is likely that these BAL quasars could be related to CSS
sources~(e.g. Kunert-Bajraszewska \& Marecki 2006). The remaining
two steep-spectrum sources, J1122+3124 and J1312+2319, exhibit
two-sided structure. According to the unification scheme, the
orientation of their jets are far from the line of sight. The
sizes of the radio sources are less than 2 kpc, the same as the
typical size of Compact Symmetric Object (CSO). However, their
central components account for more than 80\% of the total flux
density, which is significantly different from the structure of
CSOs or FR~II quasars, which have weak cores at low frequencies in
general. Given the weakness of the lobes in both objects at 18~cm,
it is plausible that their cores also have steep spectra.
Therefore, these two sources are similar to other steep-spectrum
radio BAL quasars except for the detection of weak extended lobes.
The compact, steep-spectrum central component in these two BAL
quasars suggest a link to the CSS sources.

The four flat-spectrum sources, J0724+4159, J1413+4212, J1556+3517
and J1655+3945, show core-dominated or marginal core-jet
structures in the EVN/MERLIN maps. There are two possible
interpretations for this. If their compact component is
relativisticly beamed emission from the base of the jet, the jet
in these BAL quasars is near the line of sight. Their flat spectra
are consistent with this interpretation. The degree of
polarization of a few to ten percent is also typical for such
sources. On the other hand, the flat, compact core might be
related to the GHz peaked sources (GPS)~ (e.g., Benn et al. 2005).
Noting that with an $\alpha=0$ between 3.6 cm and 20 cm for
J0724+4159, the peak frequency is likely to be around a few GHz in
the observer's rest frame. For the other three, with spectral
indexes of $\alpha=0.1$ and 0.2 between 3.6 cm and 20 cm, the
turnover frequency is also likely to be in the GHz range although
further observations are needed to confirm this.

Our results suggest that the simple unification of BAL and non-BAL
quasars by orientation is problematic. All the BAL quasars in our
sample have compact radio morphology, including two-sided
structure sources. If these BAL quasars are intrinsically small
sources with relativistic jets, they can be observed either
pole-on and edge-on. This scenario is not consistent with the
current popular disk-wind models. Basing on their radio spectrum
and morphology, the steep-spectrum BAL quasars could be classified
as CSS, while the flat-spectrum BAL quasars might be GPS
sources~(Becker et al. 2000; Gregg et al. 2000). Both CSS and GPS
sources are generally thought to represent the early stage in
quasar lifetimes ($\rm O^{'}$Dea 1998). Due to their low
luminosity, these BAL quasars might be located at the low end of
radio power. Meanwhile, as for those radio luminous BAL quasars, a
significant anti-correlation between radio loudness and the
strength of the BAL features is exhibited in a total of eleven
FR~II-BAL quasars~(Gregg, Becker \& Vries, 2006) so far. The
rarity of the FR~II-BAL quasars indicates that the period of FR-II
type combined BAL feature is very short~(Gregg et al. 2006). This
suggests an evolutionary picture in which FR~II-BAL sources are
frustrated by the obscuring BAL shroud until the quasars can boil
away enough of the material through radiation pressure. Meanwhile,
comparing the black hole mass and accretion rate between two small
samples of BAL and non-BAL quasars, Yuan \& Wills (2003) suggests
that BAL quasars have a more plentiful fuel supply than non-BAL
quasars, which might be related to the young age of this kind of
source. In addition, by using optical information from released
SDSS data, we derived black hole masses for 7 BAL quasars in our
sample within the range 9.3 $\times10^{6} \rm M_{\odot}$ to 4.5
$\times10^{8}\rm M_{\odot}$ (see detailed information in Table 4).
Despite the limited numbers in the sample, the comparatively low
black hole masses of these BAL quasars might also be connected to
their stage in the evolutionary sequence.

\section{Conclusions}
We present the results of EVN plus MERLIN polarization
observations of 8 BAL quasars at 1.6 GHz, including 4 LoBALs and 4
HiBALs with either steep or flat spectra on VLA scales. The main
conclusions are summarized as follows:
\begin{itemize}
\item{Only one steep-spectrum source, J1122+3124, shows two-sided
structure on the scale of 2~kpc. The other four steep-spectrum
sources and three flat-spectrum sources display either an
unresolved image or a core-jet structure on scales of less than
three hundred parsecs, well within the galaxy size. In all cases,
the marginally resolved core is the dominant radio component.
Making use of the phase-reference technique, celestial positions
are derived from our observations which are more accurate than
those from the VLA.}

\item{Linear polarization has been detected in the core in the
range of a few to 10 percent. Polarization, together with high
brightness temperatures (from 2$\times10^9-5\times10^{10}$K),
suggest a synchrotron origin for the radio emission. There is no
apparent difference in the radio morphologies or polarization
between low-ionization and high-ionization BAL quasars nor between
flat- and steep-spectrum sources.}

\item{We considered compact steep-spectrum or GHz peaked radio
source at the low end of radio power as the most likely
explanation for these radio BAL QSOs. Therefore, they are probably
a population of young radio sources.}

\end{itemize}

\section*{Acknowledgments}
We are grateful to I. W. A. Browne and X. W. Cao for helpful
suggestions and discussions. We thank the anonymous referee for
insightful comments and constructive suggestions. We also thank
Richard Porcas for helpful proofreading that improved the
presentation of this work. The work is supported by the NSFC under
grants 10373019, 10333020. TW acknowledges financial support NSFC
10573015.  The European VLBI Network is a joint facility of
European, Chinese, and other radio astronomy institutes funded by
their national research councils. MERLIN is operated as a National
Facility by the University of Manchester at Jodrell Bank
Observatory on behalf of the UK Particle Physics \& Astronomy
Research Council. This research has made use of the NASA/ IPAC
Extragalactic Database (NED), which is operated by the Jet
Propulsion Laboratory, California Institute of Technology, under
contract with the National Aeronautics and Space Administration.
This paper has made use of data from the SDSS. Funding for the
creation and the distribution of the SDSS Archive has been
provided by the Alfred P. Sloan Foundation, the participating
institutions, the National Aeronautics and Space Administration,
the National Science Foundation, the US Department of Energy, the
Japanese Monbukagakusho, and the Max Planck Society.


{}

\clearpage
\begin{figure}
  \begin{center}
    \mbox{
     \subfigure[J0724+4159]
     {\scalebox{0.4}{\includegraphics{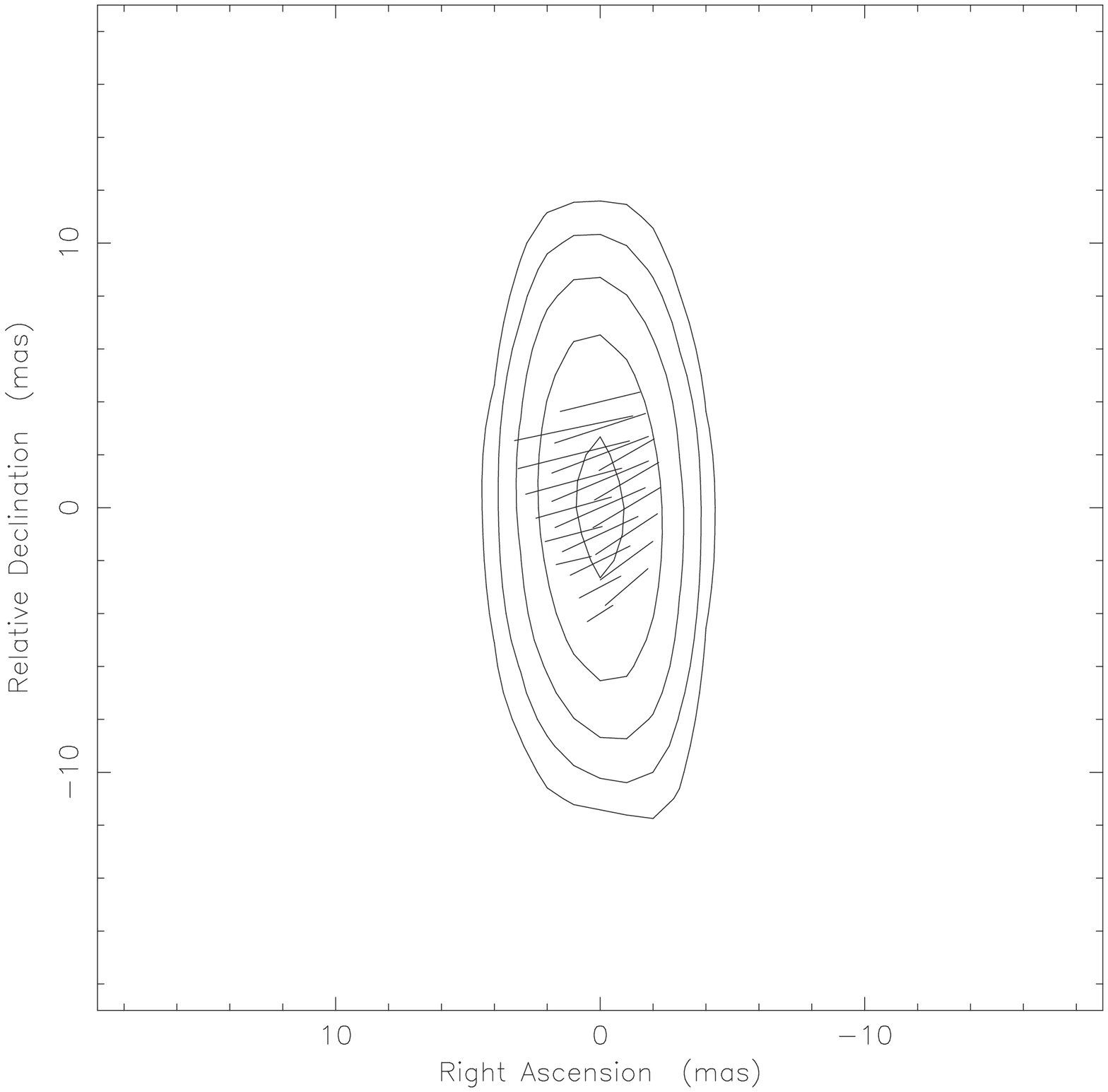}}}
      \quad

      \subfigure[J0728+4026]
      {\scalebox{0.4}{\includegraphics{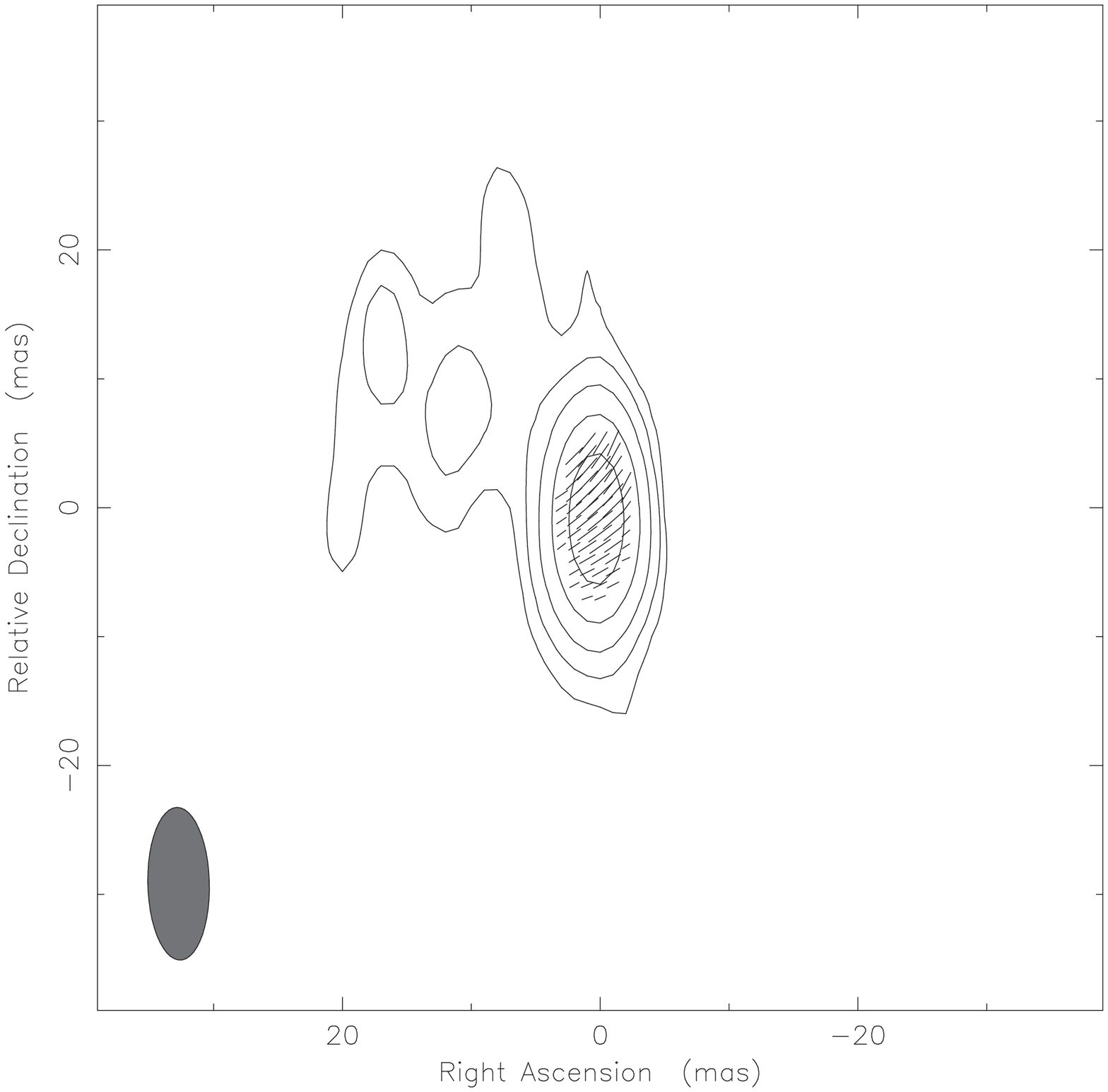}}}
      \quad}
    \mbox{
      \subfigure[J1044+3656]
      {\scalebox{0.4}{\includegraphics{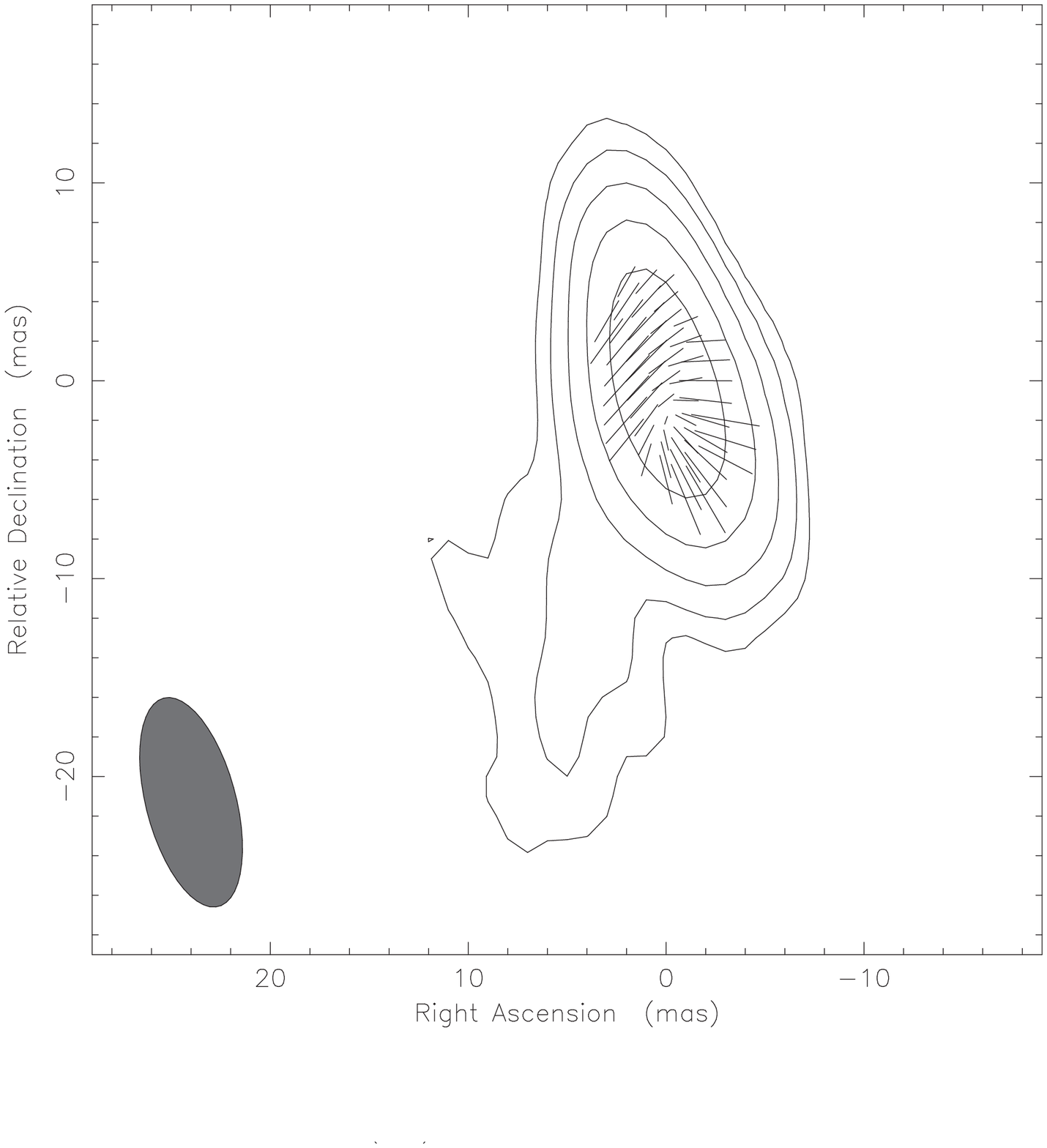}}}
      \quad

      \subfigure[J1122+3124]
      {\scalebox{0.4}{\includegraphics{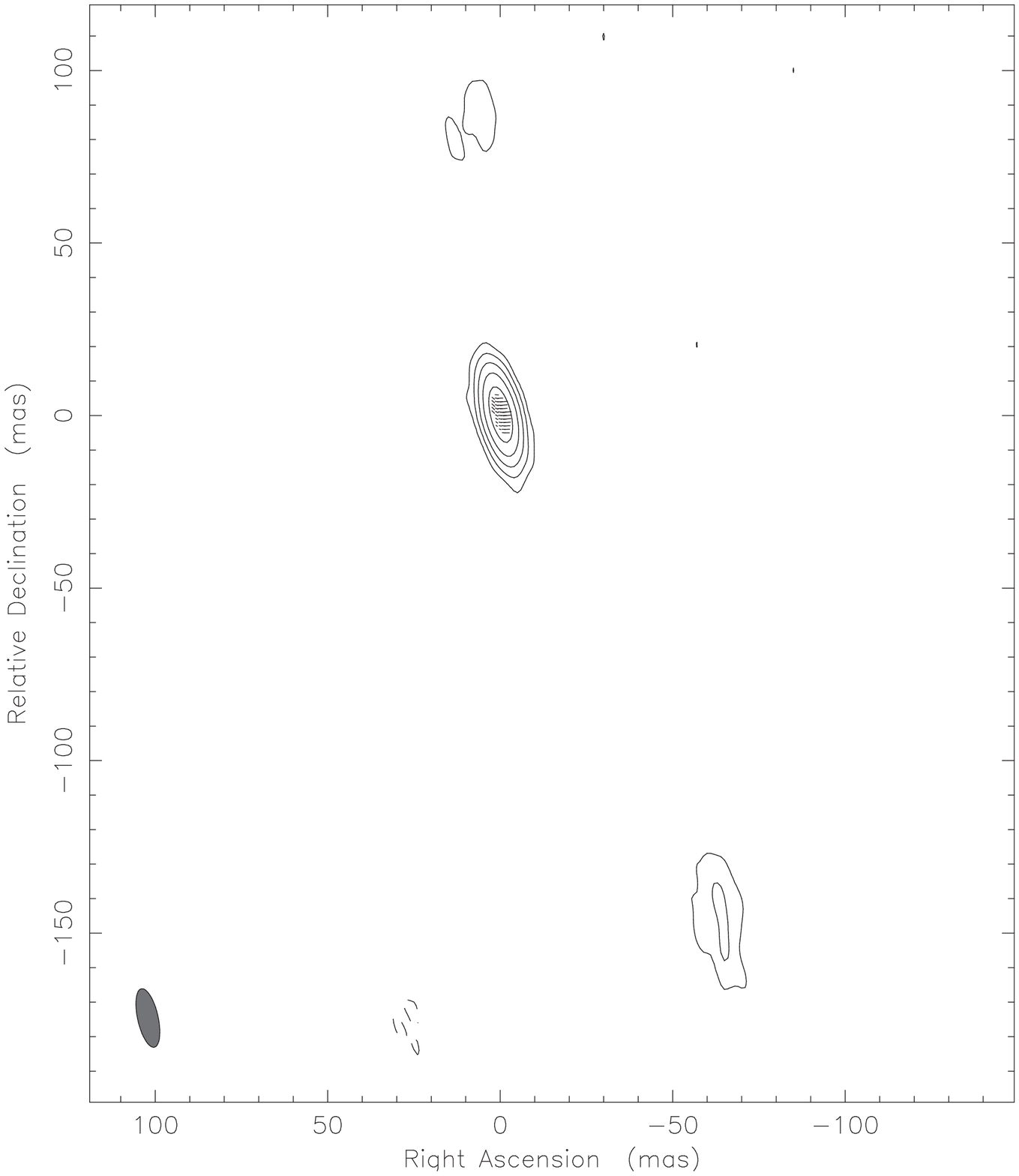}}}
      \quad}
  \end{center}
  \caption{EVN images of LoBAL quasars at 1.6 GHz.
  (a): The restoring beam is 12.2$\times$4.3 mas at P.A.=3.3$^{\circ}$,
       the contour levels are (1, 2, 4, 8, 16)$\times$0.383 mJy/beam,
       and the peak flux density is 7.0 mJy/beam.
  (b): The restoring beam is 11.8$\times$4.8 mas at P.A.=1.3$^{\circ}$,
       the contour levels are (1, 2, 4, 8, 16)$\times$0.443 mJy/beam,
       and the peak flux density is 11.2 mJy/beam.
  (c): The restoring beam is 10.9$\times$4.6 mas at P.A.=14.8$^{\circ}$,
       the contour levels are (1, 2, 4, 8, 16)$\times$0.343 mJy/beam,
       and the peak flux density is 10.8 mJy/beam.
  (d): The restoring beam is 17.4$\times$5.9 mas at P.A.=12.6$^{\circ}$,
       the contour levels are (-1, 1, 2, 4, 8, 16)$\times$0.240 mJy/beam,
       and the peak flux density is 6.71 mJy/beam.}
\end{figure}

\begin{figure}
  \begin{center}
    \includegraphics[height=.5\textheight]{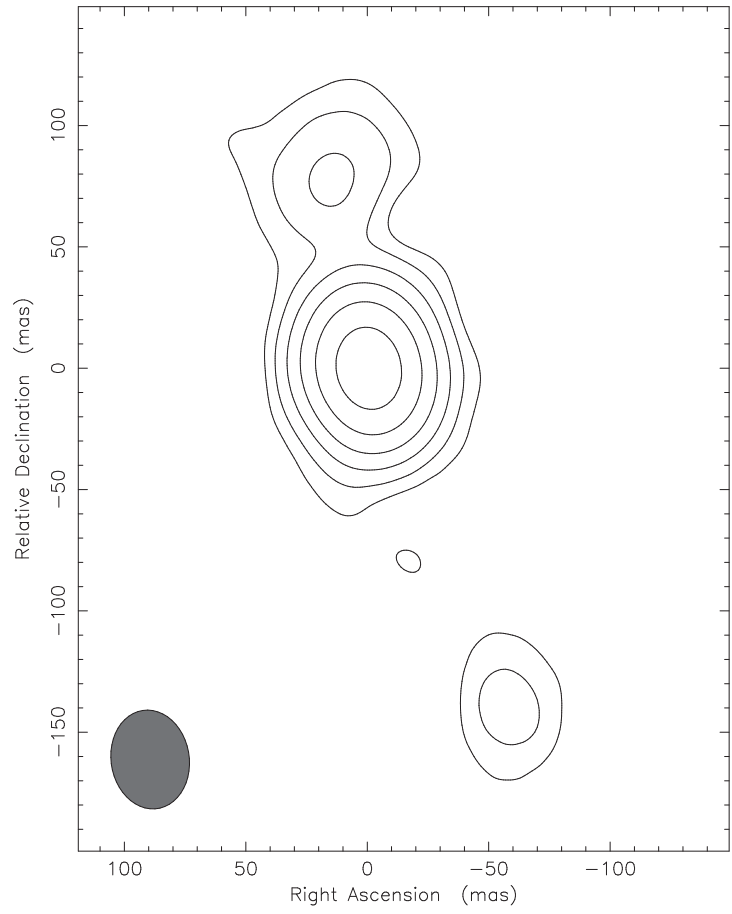}
  \end{center}
\caption{EVN+MERLIN combined image of BAL quasar J1122+3124 at 1.6
GHz. The restoring beam is 40.9$\times$ 32.2 mas at P.A. =
$8.5^{\circ}$. The contour levels are (1, 2, 4, 8, 16,
32)$~\times$ 0.167 mJy/beam. The peak flux density is 8.29
mJy/beam.}
\end{figure}

\clearpage

\begin{figure}
  \begin{center}
    \mbox{
     \subfigure[J1150+2819]{\scalebox{0.4}{\includegraphics{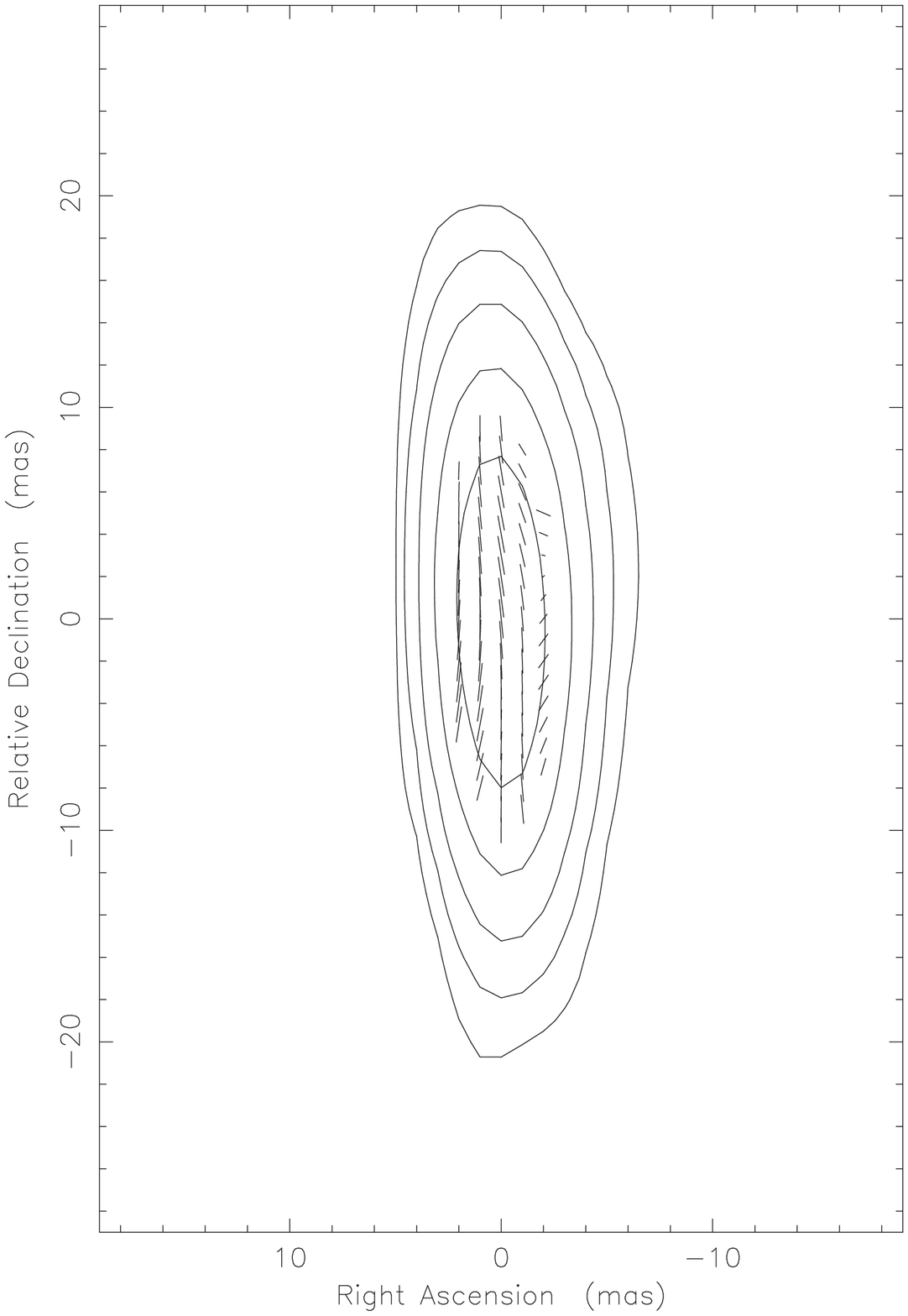}}}
      \quad

      \subfigure[J1413+4212]{\scalebox{0.4}{\includegraphics{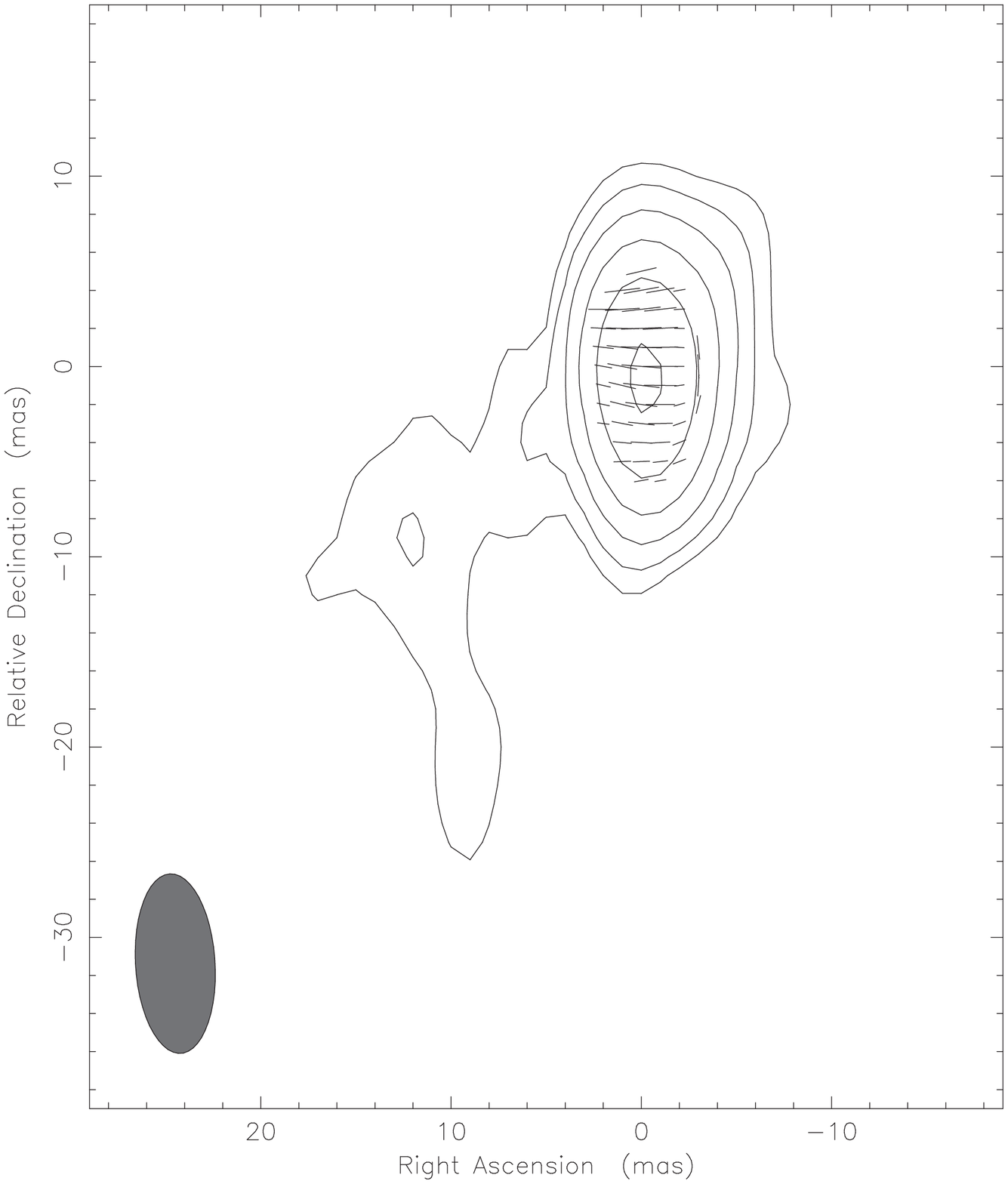}}}
      \quad}
    \mbox{
      \subfigure[J1603+3002]{\scalebox{0.4}{\includegraphics{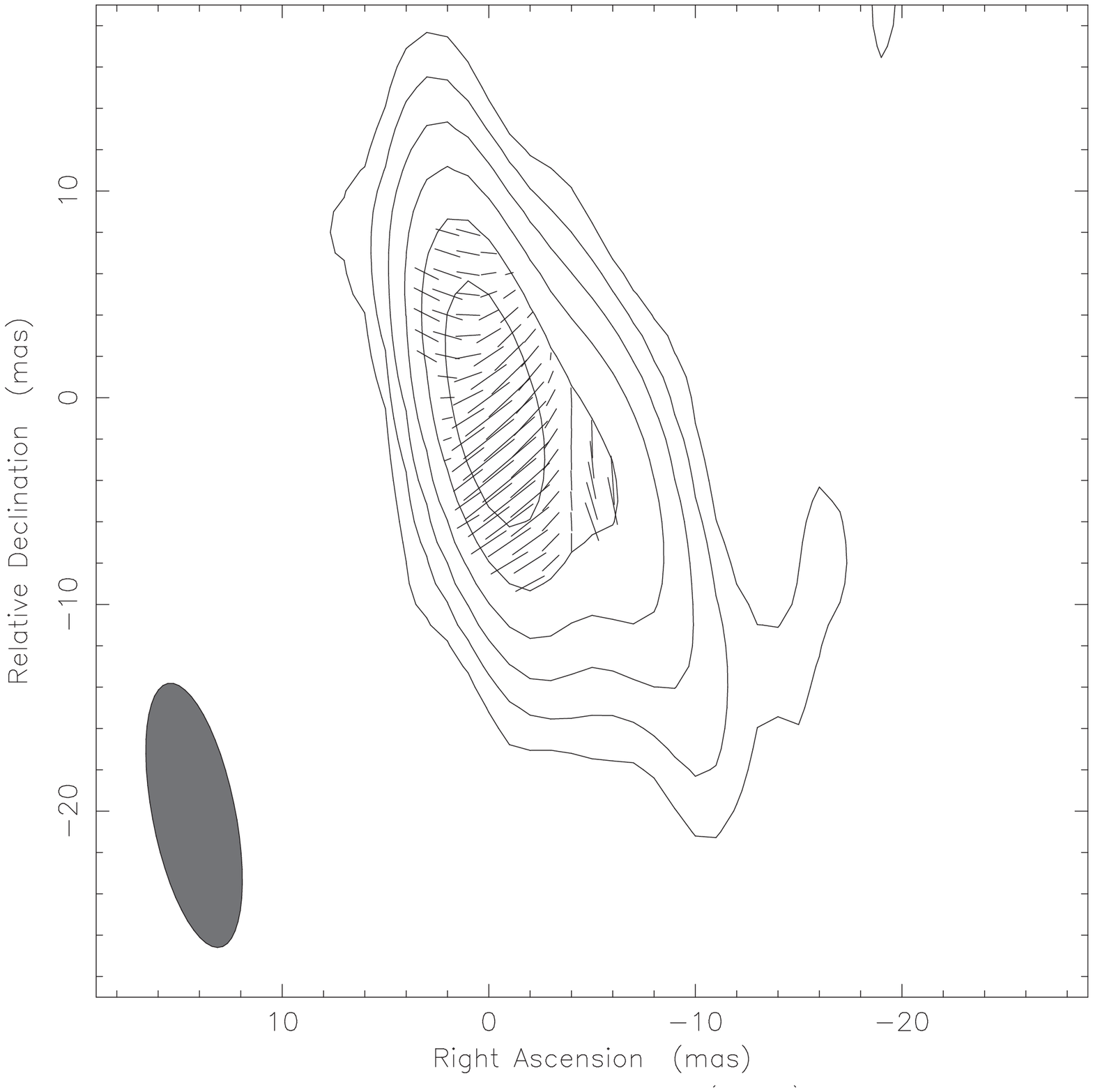}}}
      \quad

      \subfigure[J1655+3945]{\scalebox{0.4}{\includegraphics{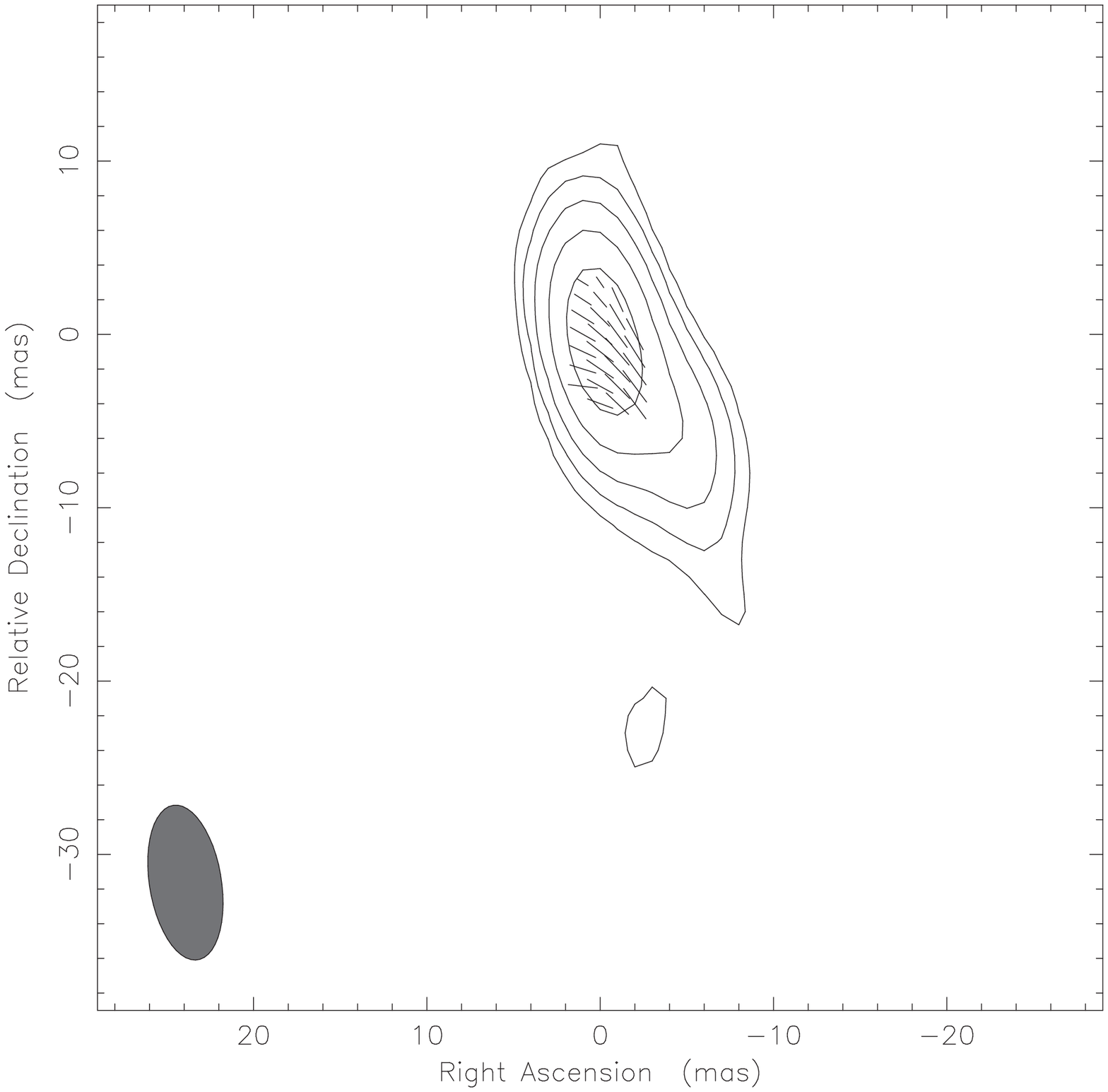}}}
      \quad}
  \end{center}
  \caption{EVN images of HiBAL quasars at 1.6 GHz.
  (a): The restoring beam is 18.1$\times$4.4 mas at P.A.=2.5$^{\circ}$,
       the contour levels are (1, 2, 4, 8, 16)$\times$0.432 mJy/beam,
       and the peak flux density is 11.7 mJy/beam.
  (b): The restoring beam is 9.5$\times$4.2 mas at P.A.=3.6$^{\circ}$,
       the contour levels are (1, 2, 4, 8, 16, 32)$\times$0.362 mJy/beam,
       and the peak flux density is 12.7 mJy/beam.
  (c): The restoring beam is 13.0$\times$4.0 mas at P.A.=11.2$^{\circ}$,
       the contour levels are (1, 2, 4, 8, 16, 32)$\times$0.669 mJy/beam,
       and the peak flux density is 37.1 mJy/beam.
  (d): The restoring beam is 9.0$\times$4.2 mas at P.A.=9.1$^{\circ}$,
       the contour levels are (1, 2, 4, 8, 16)$\times$0.313 mJy/beam,
       and the peak flux density is 8.84 mJy/beam.}
\end{figure}

\end{document}